\documentclass[acmsmall]{acmart}

\usepackage[dvipsnames]{xcolor}
\usepackage{soul}

\usepackage{graphicx} 
\usepackage{nicematrix}
\usepackage{tikz}

\usepackage{amsmath, amssymb, bbm}
\usepackage{textcomp}
\usepackage{titlesec}
\usepackage{xspace}
\usepackage{enumitem}
\usepackage{subfigure}
\usepackage{mathtools}
\usepackage{xurl}

\usepackage{appendix}
\usepackage{float}
\usepackage{sidecap}
\usepackage{pifont}
\usepackage{multicol,multirow}
\usepackage{tablefootnote}
\usepackage{algorithm}
\usepackage{algpseudocode}
\usepackage{graphicx,wrapfig,lipsum}

\newcommand{\tname}{{\sc Atlas}\xspace}
\newcommand{\iname}{{\sc BubbleTea}\xspace}

\newcommand{\fix}{\textcolor{red}}
\newcommand{\db}{\textcolor{black}}
\newcommand{\rohan}{\textcolor{black}}

\newcommand{\parab}[1]{\vspace{0.03in}\noindent{\bf #1}}

\renewcommand\footnotetextcopyrightpermission[1]{} 
\setcopyright{none}
\acmDOI{}
\acmISBN{}
\acmPrice{}

\author
{Palak}
\affiliation{
  \institution{Microsoft Research}
  \country{India} 
}
\author
{Tella Rajashekhar Reddy}
\affiliation{
  \institution{Microsoft Research}
  \country{India} 
}
\author
{Bhaskar Kataria}
\affiliation{
  \institution{Cornell University}
  \country{USA} 
}
\author
{Rohan Gandhi}
\affiliation{
  \institution{Microsoft Research}
  \country{India} 
}
\author{Karan Tandon}
\affiliation{
 \institution{Microsoft Research}
 \country{India}
}
\author{Debopam Bhattacherjee}
\affiliation{
 \institution{Microsoft Research}
 \country{India}
}
\author{Venkata N. Padmanabhan}
\affiliation{
 \institution{Microsoft Research}
 \country{India}
}

\renewcommand{\shortauthors}{Palak et al.}

\begin{document}

\newcommand{\ask}[1]{\textcolor{RawSienna}{#1}}
\newcommand{\maybe}[1]{\textcolor{RawSienna}{#1}}
\newcommand{\add}[1]{\textcolor{PineGreen}{#1}}
\newcommand{\remove}[1]{\textcolor{red}{\st{#1}}}
\newcommand{\replace}[2]{\rem{#1}\add{\ #2}}

\newcommand{\karan}[1]{\textcolor{blue}{\comment{Karan}{#1}}}

\title{Improving training time and GPU utilization in geo-distributed language model training}

\date{}

\maketitle
\sloppy 

\section*{Abstract}
The widespread adoption of language models (LMs) has caused a huge surge in demand for GPUs. Training large LMs requires tens of thousands of GPUs and housing them in the same datacenter (DC) is a challenge due to many constraints including availability of peak power. We focus on training such models across multiple DCs connected via the Wide-Area-Network (WAN). We built \tname that speeds up the training time using novel workload-aware temporal bandwidth sharing and other design choices. While \tname improves the training time, it does not completely eliminate the bubbles (idle GPU cycles). We built \iname that runs prefill-as-a-service (part of LM inference) during the bubbles thus improving the GPU utilization without any impact on training. Compared to state-of-the-art designs, \tname and \iname together achieve up to 17$\times$ faster training, and up to 94\% GPU utilization. The code will be open-sourced.

\section{Introduction}
\label{sec:intro}

\db{The advent of Generative AI and Language Models (LMs) such  as GPT\cite{gpt4:web}, Llama\cite{llama:web}, Claude\cite{claude:web} and Mistral\cite{mistral:web} has ushered in an era of widespread AI adoption across many industries from healthcare to finance to entertainment. As more sectors realize this newly unleashed potential, AI training is considered one of the larger cloud workloads of today\cite{megascale:nsdi24}.} 

Training such LMs requires a significant number of GPUs. These LMs have seen a substantial increase in the number of parameters to improve the accuracy\cite{scaling:arxiv20} and support larger number of tokens. For instance, the largest variant of the Llama model, Llama~$4$, now boasts $400$~billion parameters. Consequently, it is reported that GPT models with trillions of parameters\cite{gpt4size:web} require $10$s of thousands of GPUs\cite{gptcost:web}, and Llama models with billions of parameters require thousands of GPUs for months to train\cite{llamacost:web}.

It is desired \db{that all GPUs catering to a training workload are housed in the same data center (DC) and leverage the fast intra-DC interconnects to finish the training quickly\cite{megascale:nsdi24, metatraining:sigcomm24}.} However, 
it is \db{a non-trivial engineering challenge to deploy} a large number of GPUs in the same DC due to space, power, power density, and cooling requirements\cite{heat1:web, heat2:arxiv, heat3:hpec, heat4:begel22, heat5:web}. Such constraints, coupled with the increasing AI inferencing demand consuming up to $90\%$ of AI compute~\cite{patel2024characterizing}, are already getting reflected in user reports on not being able to provision (book) even a few GPUs\cite{geo:euromlsys24} per site. This hints at an emerging systems challenge -- getting access to thousands of GPUs at a single DC site for training LMs is hard. We here explore the intuitive yet non-trivial solution to this problem -- geo-distributing the training jobs across GPUs that span multiple DCs connected via the Wide-Area-Network (WAN). Such an approach has received attention recently\cite{geotraining:web, geotraining2:web, geotraining3:web}.





In this paper, we focus on the problem of training individual LMs on GPUs housed across multiple DCs connected via WAN. Training jobs use multiple forms of parallelism such as Data Parallelism (DP), Pipeline Parallelism (PP) and Tensor Parallelism (TP) that all have substantial communication on the critical path (\S\ref{sec:back}). Communication, needed during activation updates, gradient updates, and synchronization, incurs a significantly higher latency over inter-DC WAN than in intra-DC networks. Our experiments for doing PP across DCs show how state-of-the-art schedulers such as Varuna\cite{varuna:eurosys22} face the following limitations: (a) They end up creating bubbles (idle time; undesired) not only between the forward and backward passes in one training iteration, but also \textit{between  microbatches in the same minibatch} (details in \S\ref{sec:limit:pp}). (b) Similarly, in PP, the DCs running later pipeline stages are idle (bubble) before activations are transferred from the preceding stages. These bubbles are amplified due to slow WAN communication. Consequently, such systems achieve only $<5\%$ GPU utilization and each training iteration is severely elongated. (c) Compared to a hypothetical baseline of housing all such GPUs in a single DC, such geo-distributed training can result in an order of magnitude slower training time. More details are in \S\ref{sec:limitations}.



\db{We present \tname and \iname -- \tname addresses the key limitations in geo-distributed training to substantially improve the training time. However, it does not eliminate all bubbles. We built \iname that offers prefill-as-a-service, scheduling} the prefill phase of eligible inference requests to further reduce the bubbles. \tname and \iname \db{are independent systems that complement each other and} improve the training time by up to $17${}$\times$ compared to state-of-art and achieve GPU utilization of up to $94\%$ in cross-DC LM training.


\subsection{\tname design choices}
\label{sec:intro:tname}

\tname addresses above limitations using the following design choices.

\parab{\rohan{Simple idea with large benefits}:} \rohan{Prior works\cite{gaia:nsdi17, geo:sigcomm24, decentralized:neurips22} paint a bleak picture \db{of WAN bandwidth}. E.g., \cite{gaia:nsdi17} \db{could use} only $100$~Mbps bandwidth between nodes in Virginia and Sao Paulo on Amazon EC2. Similarly, \cite{geo:sigcomm24} found an average of $193$~Mbps between nodes in Utah and Wisconsin. Note that prior work has extensively focused on improving network designs for GPU to GPU communication on NVLink or InfiniBand\cite{nccl:web, rccl:web}, but not using Ethernet on WAN resulting in low throughput mentioned above (\S\ref{sec:limitations}). We found that PyTorch\cite{pytorch:web} simply used one TCP connection between two nodes on WAN. Our simple idea is to spawn multiple TCP connections between two nodes in the WAN to scale the bandwidth to $5$~ Gbps compared to 100s of Mbps on a single connection. This simple, intuitive idea of using multiple TCP connections substantially improves the training time.}

\parab{Coordinating pipelines:} We observe that there is no co-ordination among the DP pipelines leaving bubbles in the later stages of PP. To reduce the bubbles in later stages of PP, \db{\tname intelligently shares the WAN bandwidth across multiple pipelines and start the processing of microbatches at the later (in the pipeline) DCs sooner.} In doing so: (a) \tname improves the WAN utilization and improves the training time, and (b) interestingly, such scheduling eliminates the bubbles that arise during the microbatches.


\parab{Modeling right combination:} While the above ideas substantially improve the training time, we find that it is not always beneficial to use all GPUs. For example, imagine availability of $1000$~GPUs in one DC and $10$~GPUs in a different DC. It is better to not use the $10$~GPUs for the training, as the additional network overhead (slower WAN) results in more bubbles and elongates the training time. Conversely, in a more balanced distribution ($1000$~GPUs$\times2$), the training time could improve by distributing the job across both DCs. \tname uses a heuristic that computes the optimal split of GPUs across DCs \db{for LM training}. Additionally, it enables performance modeling, \i.e., `what-if' analysis to calculate the cost and training time for a given set of DC locations and the number of GPUs in each DC in simulation so that engineers can choose the right combinations of DCs.



\subsection{\iname}
\label{sec:intro:iname}

While \tname improves the training time and eliminates bubbles between the microbatches, it does not eliminate bubbles around forward and backward passes (\S\ref{sec:design:jct}).  Eliminating these bubbles requires either increasing the compute (by processing more layers \db{at a GPU node}) or reducing the communication \db{overhead}. However, we find that both such approaches do not work well in \db{cross-DC} training (details in \S\ref{sec:idesign}). 



To mitigate the wastage of compute during bubbles, we schedule \db{independent} workloads during the bubbles. Specifically, we find that \textit{prefill phase from the LM inference workloads} is well suited to \db{consume} such bubbles. Inference requests compose of distinct prefill \db{(digesting the prompt)} and (auto-regressive) decode phases and recent works have proposed decoupling between such phases during execution\cite{splitwise:isca24, distserve:osdi24}. The completion time of the prefill phase is highly deterministic due to known input prompt, which helps \iname schedule them effectively during bubbles in training.

\iname builds prefill-as-a-service: (a) \iname controller receives prefill requests and places them to consume bubbles in the training pipeline. (b) \iname controller determines the GPU (in the same DC) where such prefill phase runs and then transfers the KV cache to a different GPU in the same DC (outside \iname) for decode phase (following Splitwise architecture\cite{splitwise:isca24}). (c) \iname reduces the number of GPUs provisioned for inference while improving Time-To-First-Token (TTFT) for large prefills and minor penalty for small prefills. 


\subsection{Putting it together}
\label{sec:intro:all}


We implemented \tname and \iname that work hand-in-hand to simultaneously reduce the training time and improve GPU utilization \db{in cross-DC LM training}. Through testbed experiments and large-scale simulations, we show that: (a) \tname can reduce the training time up to $17${}$\times$ compared to state-of-art mechanisms including Varuna\cite{varuna:eurosys22}. (b) While significant benefits can be attributed to the intuitive idea of using multiple TCP connections, the other idea of intelligently sharing the WAN bandwidth improves the training time by up to $1.82${}$\times$ compared to state-of-art schedulers. (c) \tname architecture is effective in using PP across DCs. Such an architecture can scale the throughput with more DCs. (d) Algorithm in \tname to calculate the optimal number of GPUs per DC is effective and fast. (e) \tname and \iname co-ordinate effectively, and \iname is able to achieve the GPU utilization up to 94\%.

This work does not raise any ethical issues.

\section{Background}
\label{sec:back}


Most of the LMs such as GPT\cite{gpt4:web}, LLama\cite{llama:web}, Claude\cite{claude:web}, Mistral\cite{mistral:web} use transformer based models\cite{transformer:neurips17} with many layers. Each transformer block comprises various components, including attention mechanisms and feedforward neural networks (FFNs), each containing its own neural network. A training job learns the parameters of neural networks.

In each training iteration, a model takes a few samples of a data called \textit{minibatch} and performs a \textit{forward pass} that computes the loss values for the data samples followed by a \textit{backward pass} that computes the gradients. Finally, the model parameters are learnt by applying the negative of the gradients. Recent works save GPU memory by re-running the forward pass just before the backward pass\cite{varuna:eurosys22}, called \textit{recomputation}. Due to the massive sizes of the models (billions to trillions of parameters), the training job is distributed across multiple GPUs. The following parallelism modes distribute the job:

\parab{Data parallelism (DP):} In this mode, the model (or a subset of its layers) is replicated across GPUs and different minibatches are fed to such replicas. At the end of the iteration (one forward and backward pass), gradients are averaged through \textit{all-reduce}\cite{allreduce:pdc09} where the communication between replicas is on the critical path. 


\parab{Pipeline parallelism (PP):} While DP helps in speeding up the training time, PP helps in fitting larger models across GPUs. In PP, different layers of the model are assigned to different GPUs. One GPU sends the activations (in forward pass) to the next GPU over network. Similarly, in the backward pass, the gradients are sent between GPUs over network. Furthermore, the minibatch is further split across different \textit{microbatches} that are pipelined in execution. Recent works try to overlap compute of one microbatch with communication of a different microbatch to improve training time\cite{gpipe:neurips19, pipedream:sosp19, varuna:eurosys22, megatron:arxiv19}. \db{The critical path is shaped by the slowest (due to slower communication, computation, or both) pipeline stage.}

\parab{Tensor parallelism (TP):} Like PP, TP helps in fitting models across GPUs. Unlike PP, TP splits individual layers across different GPUs and uses all-reduce for communication. \db{TP requires significantly higher network bandwidth than DP and PP due to frequent synchronization across shards\cite{tp:arxiv25}.}

\parab{Expert parallelism (EP):} In the mixture-of-expert (MoE) models~\cite{gshard:arxiv20, switchtransformer:22}, the data samples are routed to one or a few \textit{experts}. In this work we focus on the \textit{dense} models (e.g., Llama 3.1) that do not use experts. The principles from this paper apply to MoE models (discussed in Appendix \S\ref{sec:app:moe}).

\parab{Sequence parallelism (SP):} In PP, we scale sequence dimension and is usually applied at the end of the pre-training to increase model context window\cite{sp:arxiv23, sp2:arxiv23}.

\parab{Multiple forms of parallelism:} Training large models usually 
\db{involves all forms of parallelism described} above. However, the key is determining how to split the job across such different forms of parallelism. 

\subsection{Geo-distributed training}
\label{sec:back:geo}

Training large language models (with billions of parameters) necessitates a substantial number of GPUs to minimize training latency. Many previous works assume that all GPUs used for training are located within the same datacenter (DC)\cite{alpa:osdi22, gpipe:neurips19, varuna:eurosys22, megatron:arxiv19, antman:osdi20, cassini:nsdi24, sia:sosp23, crux:sigcomm24}. These works benefit from the favorable network conditions within a single DC, such as low latency and high bandwidth. However, consolidating all GPUs in a single DC is becoming increasingly challenging \db{as ($1$) many GPUs (up to $90\%$ of all compute cycles~\cite{patel2024characterizing}) are being allocated to inference workloads\cite{chatgptgpus:web} and ($2$) DCs hit the power draw and cooling thresholds due to high GPU power density\cite{dynamollm:hpca25}.} 

\begin{wrapfigure}{l}{0.6\textwidth}
\centering
\includegraphics[width=0.5\textwidth]{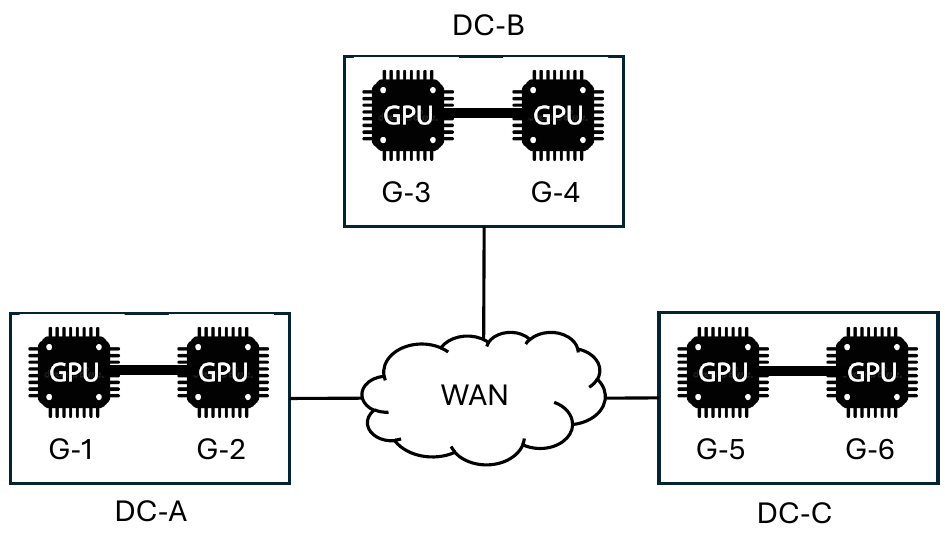}
\caption{Experimental setup consisting of 6 GPUs in 3 DCs connected by WAN. WAN offers lower bandwidth compared to intra-DC bandwidth (thicker lines).}
\label{fig:limit:setup}
\end{wrapfigure}

A potential solution is to distribute the training across GPUs located in different DCs, which are connected via WANs\cite{swan:sigcomm13, b4:sigcomm13}. Unfortunately, WANs typically do not offer the same network capacities as intra-DC networks. Most WAN links (connecting DCs) range from a few hundred Gbps to a few Tbps\cite{swan:sigcomm13, b4:sigcomm13, b4after:sigcomm18}, and DCs in different countries may experience latencies from tens to hundreds of milliseconds \db{(versus sub-millisecond intra-DC latency)}.

\db{Private WANs have seen a surge in traffic demand in recent years\cite{swan:sigcomm13, b4:sigcomm13, wanincrease:web}.}
For example, Meta reported $13${}$\times$ increase in WAN traffic in $6$~years (not including cross DC training traffic). However, it is challenging to increase the WAN capacity as laying out fiber and (in some cases) under-sea cables can take years~\cite{wanincrease:web}. Given this provisioning challenge and the increasing demand, WAN bandwidth is a constrained resource.

However, TP, EP and SP operate within layers and require substantially higher network bandwidth and small latency (that's offered by NVLink or InfiniBand)\cite{tp:arxiv25, gshard:arxiv20}. As mentioned above, WAN suffers from high latency and low bandwidth compared to networks within DCs. On the other hand, DP and PP require much smaller bandwidth and are good candidates for training over WAN.

\section{Limitations of existing designs}
\label{sec:limitations}

\db{It is critical to understand the impact of WAN latency and capacity constraints on LM training workloads that span multiple DC sites. In this section, we quantify the WAN communication overhead on the different modes of parallelism.}


\begin{table}
\centering 
\caption{Bandwidth for single TCP connection for different WAN latencies.}
\label{tab:mot:bw}
{
\small
\begin{tabular} {|c|c|c|c|c|}
\hline
Latency (msec) & 10 & 20 & 30 & 40 \\
\hline
Bandwidth (Mbps) & 1220 & 600 & 396 & 293 \\
\hline
\end{tabular}
}
\vspace{-0.1in}
\end{table}

\parab{Setup:} We spawn $6$ A100 GPU nodes across $3$ DCs ($\times${}$2$ GPU nodes each). Each node has a single A100 GPU. \db{For controlled experiments, we emulate the WAN latency using \texttt{tc}.}
Fig.\ref{fig:limit:setup} shows the experimental setup. \db{We use PyTorch framework for training.}
We measure the training time of iterations as we vary the WAN latencies. Lower (higher) WAN latencies indicate that the DCs are closer (farther).

\parab{Baseline models:} We use two baseline models: (a) GPT-A (similar to GPT-3) with context length (L) of 4K and hidden dimension (H) of 4K, each layer is 412 million parameters, (b) GPT-B (bigger model than GPT-3) with L,H = 6K,8K, each layer is 1.2 billion parameters. We limit the number of layers to fit on 6 GPUs. 

Note that the individual layer sizes for GPT-B are higher than Llama 3-70B model (such a model has roughly 80 layers resulting in each layer with size of roughly 875 million parameters). We use a smaller number of layers to fit the model on 6 GPUs.


\subsection{Impact of WAN latency on DP}
\label{sec:limit:dp}

\begin{figure*}
\centering
    \begin{minipage}{.45\textwidth}
        \centering
        \includegraphics[width=0.95\textwidth, page=1]{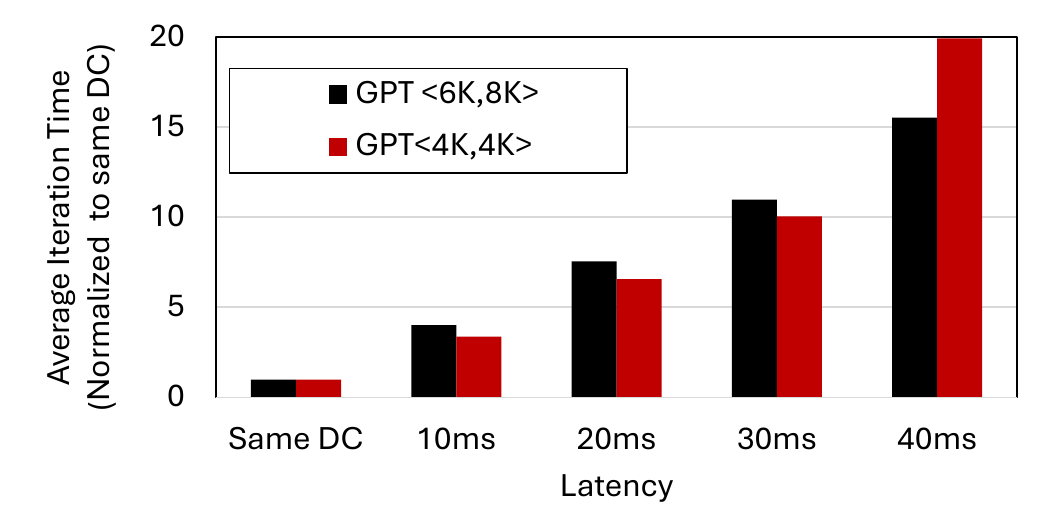}
        \caption{Slow-down in training time in DP as we increase the WAN latency. \db{$y$-values are multipliers and hence unit-less.}}
        \label{fig:limit:dp}
    \end{minipage}%
    \hspace{0.2cm}
    \begin{minipage}{0.45\textwidth}
        \centering
        \includegraphics[width=0.95\textwidth, page=2]{figs/mot-dp-pp.pdf}
        \caption{Slow-down in training time in PP as we increase the WAN latency. \db{$y$-values are multipliers and hence unit-less.}}
        \label{fig:limit:pp}
    \end{minipage}%
\end{figure*}

In this experiment, all the nodes form a single DP ring, i.e., the model is replicated across all the nodes in the ring. Individual nodes are fed different minibatches, and the gradients are averaged out during the all-reduce communication phase.

We measure the impact on the training time due to increasing WAN latencies. Fig.\ref{fig:limit:dp} shows the training time for individual iterations. \db{While compute time stays the same, increase in WAN latency affects the all-reduce time. Now, as shown in Table \ref{tab:mot:bw}, higher WAN latency translates to lower achievable bandwidth and, hence, higher iteration time.}


We make the following observations: (a) As shown in Fig.\ref{fig:limit:dp}, there is a significant slow-down in the training time as the WAN latency increases. For a WAN latency of $40$~msec, the training time slows down by more than $15${}$\times$ \db{compared to the same DC baseline}. \rohan{(b) This slowdown is mainly because communication latency increased drastically\footnote{Analytically, for a model with $P$ parameters, $N$ nodes in the all-reduce ring, and $BW$ bandwidth, communication time per iteration can be calculated as $\frac{4 \cdot P \cdot (N-1)}{N \cdot BW}$ (extra factor 2 is due to fp$16$).}. For WAN latency of $40$~msec, $93\%$ and $95\%$ of the iteration time was spent in communication for the two models}. (c) We find that all-reduce takes major chunk of the training time. For latency = $40$~msec for GPT-B, all-reduce consumed roughly $98\%$ of the training time. (d) PyTorch only uses one TCP connection between one pair of nodes for all-reduce. As the WAN latencies go up, the bandwidth is reduced (Table \ref{tab:mot:bw}) that in turn increases the training time. (e) Lastly, WAN consists of TCP based interconnections where the data to be communicated is first copied to CPU that runs the TCP connection for communication. That said, time to copy the data between GPU-CPU is minuscule ($<${}$1\%$) and majority of the time is consumed due to communication over the WAN. 



\subsection{Impact of WAN latency on PP}
\label{sec:limit:pp}

\begin{figure*}[t]
\centering
\includegraphics[width=0.95\textwidth, page=1]{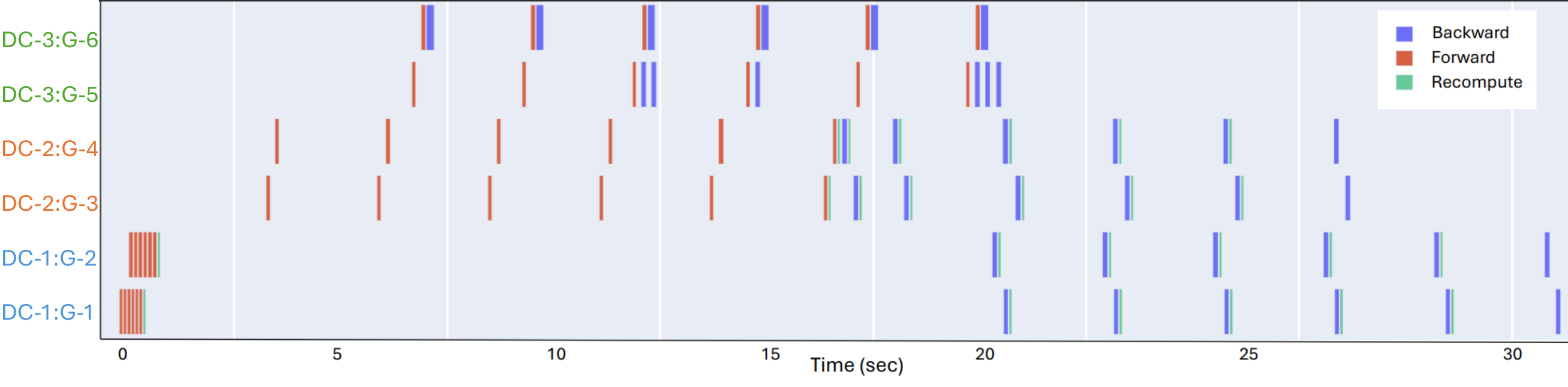}
\caption{Execution of compute phases (forward, backward, recompute) in PP in Varuna across 6 GPUs (G-1 to G-6) in 3 DCs for L,H = 6K,8K and WAN latency of 40 msec.}
\vspace{-0.1in}
\label{fig:limit:pptime}
\end{figure*}

We evaluate the impact of WAN latencies on the training time when the nodes run PP. In this experiment, we split the number of layers across all 6 nodes. The nodes in the same DC get adjoining layers to minimize the cross-DC traffic (and improve training time). Note that the existing schedulers including GPipe\cite{gpipe:neurips19} and Varuna\cite{varuna:eurosys22} try to overlap compute with communication to reduce the job completion time. Fig.\ref{fig:limit:pp} shows the increase in the training time as we increase the WAN latencies in Varuna scheduler that optimizes the PP scheduling over GPipe and Megatron\cite{megatron:arxiv19}.

Like DP, PP also shows significant slowdown in the training time \db{as WAN latency increases resulting in a reduction in the achievable bandwidth}. 
We also observe that the slowdown in PP is smaller than the slowdown in DP ($y$-range different for Fig.~\ref{fig:limit:dp} \& \ref{fig:limit:pp}) at the same WAN latency. 


Fig.\ref{fig:limit:pptime} shows the timeline of execution of PP for different microbatches for their forward pass, backward pass, and recomputation in Varuna for WAN latency = 40 msec. The training job starts in GPU G-$1$ and all microbatches execute back-to-back. The activations are sent to G-$2$, which is very fast as both the GPUs are in the same DC. However, as G-$3$ is in a different DC, the activations from G-$2$ to G-$3$ are sent over the low-bandwidth WAN. As a result, the activations take significant time (around $2.5$~seconds) to transfer from G-$2$ to G-$3$\footnote{The size of activations (and gradients) is $B \cdot L \cdot H$ (B = batch size, L = seq. length, H = hidden size) }. The delay continues to be observed for subsequent microbatches as well as for the next GPUs (G-$3$ to G-$6$). GPU G-$6$ processes the backward pass at the end of each forward pass and starts sending the gradients for the other GPUs in reverse order (G-$6$ to G-$1$). Like activations, gradients also suffer due to lower WAN bandwidth. 


We make the following observations: (a) Even with a state-of-the art scheduler like Varuna, we see significant bubbles between the forward and backward passes and \textit{also between microbatches}. \rohan{(b) This slowdown is mainly because communication latency increased drastically. For WAN latency of $40$~msec, $89.8\%$ and $91\%$ of iteration time was spent in communication for the two models}. (c) Nodes G-$5$ and G-$6$ are idle for substantial time in the beginning when the activations are transferred between G-$2$ to G-$4$. (d) Like DP, PP also used one TCP connection between individual pairs of nodes \db{when run with PyTorch}. (e) Bandwidth is \textit{not} shared across activations and gradients in PyTorch. When activations for one microbatch are transferred between two GPUs, activations for subsequent microbatches are queued (instead of parallel transfers). This helps finish transfer of existing microbatches and unblock the next GPUs for further processing. Lastly, as activations and gradients are sent in opposite directions, they do not compete for the same WAN bandwidth.

\subsection{Impact of WAN latency on TP}
\label{sec:limit:tp}
TP requires high performant networks with low latency and very high bandwidth (BW) due to all-reduce phase on  the critical path. It is mostly advised to run TP within a node (with multiple GPUs)\cite{tesseract:icpp22, tp:ispass24} where GPUs are connected via NVLink offering more than 800 Gbps BW. WAN networks do not provide such high BWs and we do not run TP over WAN. E.g.,  even with 40 msec WAN latency, the BW is just 293 Mbps (Table \ref{tab:mot:bw}) (orders of magnitude smaller than NVLink).

\section{\tname}
\label{sec:design}

\db{We quantified in the previous section how}
lower bandwidth on the WAN substantially elongates \db{training time for} different modes of parallelism. We built \tname toward improving the training time. We touched upon the summary of design choices in \S\ref{sec:intro:tname}. We expand on them next.

\subsection{Using multiple TCP connections}
\label{sec:design:tcp}

As mentioned in the previous section, PyTorch\cite{pytorch:web} simply uses one TCP connection for communication. As shown in Table \ref{tab:mot:bw}, as latency between the nodes increases, the \db{achievable} TCP bandwidth between the nodes shrinks. \rohan{Recently high bandwidth mechanisms such as NVLink, InfiniBand and RDMA have been improved for longer distances\cite{rdmadist:web}. However, they do not work beyond few hundreds of miles\cite{rdmadist:web}. Most of the cross-region DCs easily fall beyond such distances.}  

\begin{wrapfigure}{l}{0.55\textwidth}
\centering
\includegraphics[width=0.52\textwidth, page=2]{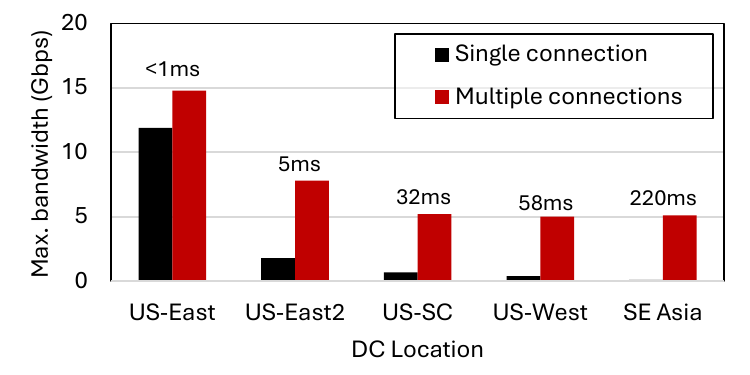}
\caption{Bandwidth for single and multiple TCP connections. The server is located in US-East, and we vary the location of the client (X-axis). US-SC denotes US South-central DC. The numbers over the bars denote one-way latencies.}
\vspace{-0.2in}
\label{fig:design:tcp}
\end{wrapfigure}

\db{\tname enables the use of multiple parallel TCP connections between VMs in different DCs to scale the bandwidth.} Fig.\ref{fig:design:tcp} shows the 
\db{achievable} bandwidth for single and multiple TCP connections as we vary the distance (hence, WAN latency) between DCs. We have a server VM in Azure US-East DC. To vary the WAN latency, we select client VMs in different DCs in US and Asia. We measure the bandwidth using \texttt{iperf3} and Cubic as TCP transport protocol. The $x$-axis shows the DC locations for clients. All VMs use Azure \rohan{F$32$as\_v$6$ series\cite{azurevm:web} with $32$ cores and $20$~Gbps NICs}. We also tested with  BBR\cite{bbr:comm17} (delay-based congestion control) flows and observed similar trends as discussed below. We set (multiple) TCP connections to achieve maximum bandwidth.




As we spawn more concurrent TCP connections, the cumulative bandwidth grows. Interestingly, we observe that such cumulative bandwidth 
\db{does not drop below} $5$~Gbps even when the nodes are at a considerable distance, and does not increase even if we increase the number of concurrent connections. We believe that this is due to the rate-limiting at the hypervisor. AWS also throttles the bandwidth at $5$~Gbps\cite{skyplane:osdi23}. That said, the key take-away is that, \tname can help get up to $5$~Gbps 
between two nodes on WAN irrespective of their distance. As a result, \tname can cut down the 
data transfer latency substantially. E.g., instead of using $250$~Mbps on a single TCP connection, now \tname can get $5$~Gbps over multiple connections -- cutting down data transfer latency by $20${}$\times$. \tname uses all available bandwidth (e.g., 5 Gbps) between two nodes \rohan{by spawning multiple TCP connections. We profile the bandwidth for single TCP connection between nodes in different DCs and set the number of TCP connections to consume 5 Gbps between nodes.}



We believe that the $5$ Gbps limit between two nodes is not unreasonable. The WAN link capacity between two WAN routers typically ranges from a few hundred Gbps to a few Tbps\cite{radwan:sigcomm18} and 
\db{even $100$ nodes at full-throttle could consume 500 Gbps (large fraction of total capacity) -- things add up quickly at scale. \tname helps consume the available WAN capacity efficiently while not violating any cloud-provider specific rate-limiting enforcement.}

Lastly, we leave exploring MPTCP for future work. Currently, we prefer using multiple TCP connections over other alternatives such as MPTCP due to: (a) cloud providers are enforcing bandwidth caps for individual hosts; spawning multiple connections increases the aggregate bandwidth per host linearly with number of TCP connections and are able to reach the bandwidth limit (enforced by cloud provider), (b) MPTCP is primarily used for leveraging multiple interfaces; in our settings, each node has a single NIC, (c) multiple TCP connections are supported by all cloud offerings out of box.



\subsection{Multiple forms of parallelism across DCs}
\label{sec:design:3d}

Recall that, we use multiple forms of parallelism to speed up training (\S\ref{sec:back}). The key question is how to divide GPUs in different DCs for such forms of parallelism. As mentioned in \S\ref{sec:limitations}, TP, EP, SP requires significant bandwidth. Thus, \tname uses them only across intra-DC nodes connected via NVLink (in the same node), RDMA or Infiniband. This leaves \tname with deciding the division of GPUs across DCs for PP and/or DP.

\parab{PP across DCs is beneficial:} Prior works including \cite{geo:euromlsys24} have shown that doing PP across DCs and DP within DCs is beneficial in geo-distributed training. Recall that PP tries to overlap compute with communication. Bubbles in compute are formed when communication takes longer than compute. For context length $L$, hidden size $H$, and batch size $B$, communication time is O($BLH$) (linear with L and H), while the compute time is O($BLH^2$) (for MLP layers) + O($BHL^2$) (for attention layers) -- quadratic with L and H\cite{megatron:arxiv19}. Thus, the gap between compute and communication shrinks as model size grows (increasing L, H, or both). On the other hand, we can do DP across DCs where nodes in the same all-reduce ring are in different DCs. However, the all-reduce runs on critical path and its time depends on number of parameters of a layer (that in turn depends on H). It only increases as model size increases. Thus, we choose to do PP across DCs and DP within DCs. 

\begin{figure*}[t]
\centering
\includegraphics[width=0.95\textwidth, page=1]{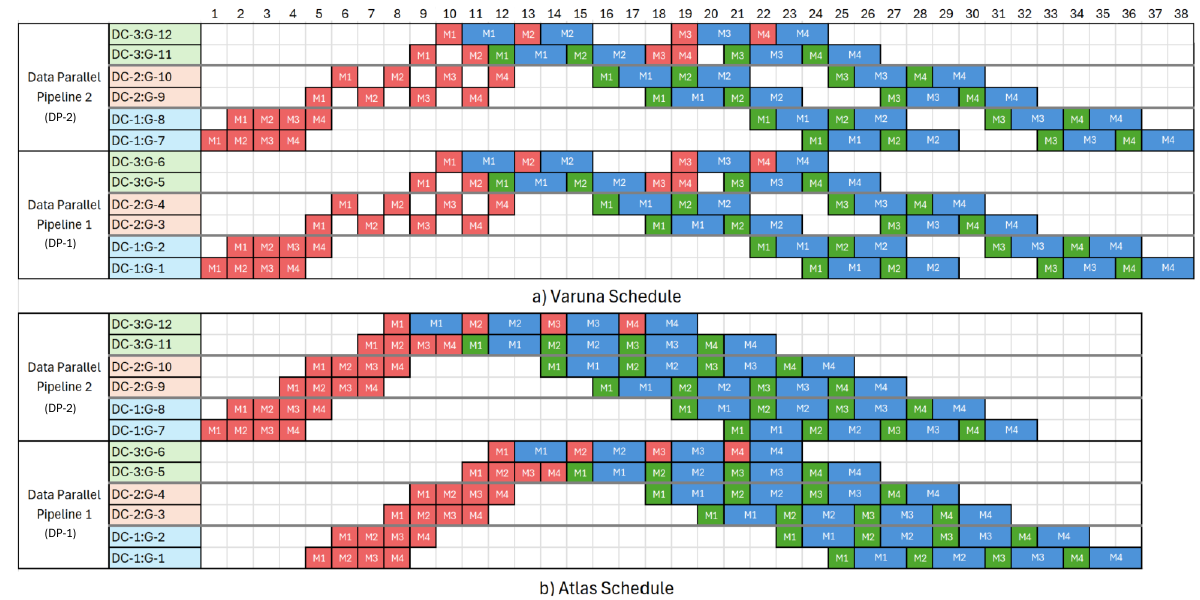}
\caption{Bandwidth sharing across DP pipelines in (b) \tname versus (a) Varuna. \db{$x$-tics represent time-slots}. M1 to M4 denote microbatches. Red, Green and Blue boxes indicate forward, recompute and backward passes.}
\vspace{-0.1in}
\label{fig:design:dpsharing}
\end{figure*}

We structure DP+PP parallelism as shown in Fig.\ref{fig:design:dpsharing}(a) where we use the structure using Varuna (we improve it as detailed in the next section): (a) We have multiple DP pipelines to scale throughput. (b) Each DP pipeline contains PP where subset of layers are assigned to individual GPUs. The PP runs across GPUs in different DCs. For example, let's consider $6$ layers. In DP-1, PP is assigned across $6$ GPUs in $3$ DCs, each GPU is assigned one layer. (c) For different DP pipelines, each layer is assigned in the same DC. For example, layer-$1$ is assigned to GPUs G-1 and G-7 (in DC-1). Thus, the all-reduce ring (that runs for each layer) only runs across nodes in the same DC. (d) TP, EP, SP is assigned across GPUs on the same node (or nodes in the same DC).  

\subsection{Improving training time through co-ordination}
\label{sec:design:jct}

When running multiform parallelism according to the above architecture, the existing schedulers such as Varuna and Gpipe assume there is no co-ordination between different DP pipelines. As a result, each  DP pipeline gets its own share of bandwidth (e.g., $5$~Gbps) and roughly follow the same schedule across different pipelines as shown in Fig.\ref{fig:design:dpsharing}(a). In essence, such sharing is \textit{spatial}, as each pipeline has its own share of bandwidth.

Note that the nodes higher in the PP pipeline (e.g., node G-11 and G-12 in Fig.\ref{fig:design:dpsharing}(a)) are idle early in the execution as they have not received their microbatches. The delay is exasperated due to slower data transfer over WAN. We observe that there is higher aggregate bandwidth available by using multiple nodes across DP pipelines. In \tname, different DP pipelines work in tandem -- \rohan{first, we split the data of one DP pipeline across multiple nodes from other DP pipelines not using their bandwidth at that time. This way, each node gets 5 Gbps across WAN, but we get higher bandwidth using multiple nodes in parallel. To do so, the node that wants to send the data over WAN, first distributes the data to other nodes in different DP pipelines using higher intra-DC bandwidth (100+ Gbps). Next, all such nodes across multiple DP pipelines send the data across WAN in parallel. This way, the DP pipeline can speed up sending its activations to the next nodes in other DCs. We call it \textit{temporal bandwidth sharing}.}



We illustrate this key idea in Fig.\ref{fig:design:dpsharing}(b). Imagine ratio of communication to compute latency (C) to 2$\times$. 
\db{Imagine there are $2$ DP pipelines each leveraging $5$~Gbps WAN bandwidth between two nodes in different DCs. With schedulers like Varuna, as in Fig.\ref{fig:design:dpsharing}(a), downstream GPUs in the pipeline (PP) in a different DC (node G-11) need to wait for longer for activation transfer to happen using $5$~Gbps WAN bandwidth. with \tname, the entire $2\times5=10$~Gbps bandwidth (across both DP instances) is available to each PP thus speeding up activation transfers to just $1$ time-slot instead of $2$.}
As a result, first DP instance progresses faster and makes the microbatches available to node G-11 and G-12 in DC-$3$ quickly, which then has \textit{cascading effects} to process next batches sooner. These nodes then start their backward pass sooner too. Cumulatively, such temporal sharing improves the training time \db{(Fig.\ref{fig:design:dpsharing}(b),  $x$-range is shorter ending at time-slot $36$ instead of $38$).} \S\ref{sec:eval:tempo} shows that such a co-ordination based scheduler can improve the training time up to 1.52$\times$ compared to Varuna.

\parab{Bubble consolidation:} The second key benefit of temporal bandwidth sharing is that it reduces the bubbles between microbatches. Notice the bubbles at node G-9, time-slot $5$-$11$ (Fig.\ref{fig:design:dpsharing}(a)) -- these bubbles are formed due to C $>$ 1 (communication taking longer than compute). \rohan{Using temporal bandwidth sharing, we can remove such bubbles between microbatches by setting the number of DP pipelines to C. This way communication takes same time as compute, removing all such bubbles between microbatches\footnote{If compute and communication take 10 and 30msec, then by using 3 DP pipelines, we reduce communication to 10msec.}. In other words, we can \db{consolidate} bubbles together by using temporal bandwidth sharing.} \db{This consolidation leads to continuous bubbles} and more opportunities to schedule other workloads during these bubbles as discussed in \S\ref{sec:idesign}.

\begin{wrapfigure}{l}{0.5\textwidth} 
\centering
\includegraphics[width=0.45\textwidth, page=1]{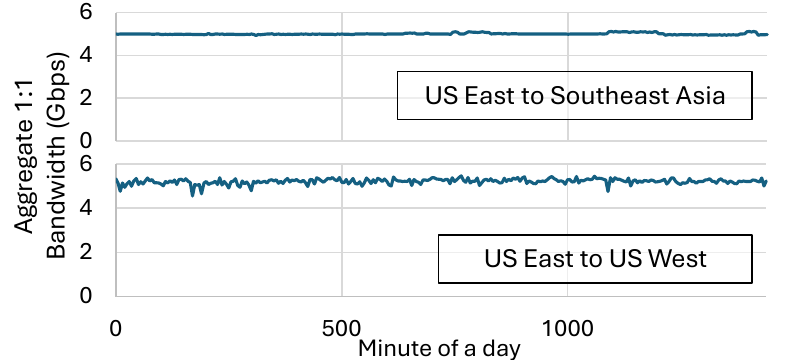}
\caption{Fluctuations in WAN bandwidth.}
\vspace{-0.1in}
\label{fig:design:wan}
\end{wrapfigure}


\parab{Addressing jitters and bandwidth fluctuations, stragglers:} 
We measure the bandwidth for $24$ hours from a VM in US-East DC (node A) to VMs in Southeast Asia (node B) and US-West (node C) DCs in Azure. We found that the bandwidth variations are quite small (Fig.\ref{fig:design:wan}). Surprisingly, although the distance between nodes A-B is larger, the bandwidth variation is much smaller where coefficient of variance (CoV) is just $0.8\%$ (CoV for A-C is $2.3\%$). This is because WANs are well provisioned, have multiple mechanisms for masking network failures\cite{swan:sigcomm13, b4after:sigcomm18} and suffer from relatively small packet loss~\cite{saving:conext24}. As such fluctuations are small, \tname tries to eliminate bubbles altogether thus ramping up (costly) GPU utilization. Even if there are stragglers, bubbles around microbatches serve as a cushion. For GPU failures, we use the existing checkpointing mechanisms such as asynchronous methods and in-memory approaches\cite{check:eurosys24}. That said, improving checkpointing over WAN is future work.

Lastly, we do not advocate training models using Internet as a transport as Internet suffers substantially higher packet losses\cite{saving:conext24}.

\subsection{Schedule for temporal bandwidth sharing}
\label{sec:design:schedule}

\tname uses a heuristic that calculates the schedule for forward and backward passes, as detailed next. The schedule is pre-computed by \tname before the training starts, and adjusts to any stragglers.

\parab{($1$)} \db{\tname groups DP instances into \textit{DP-cells} during the initialization phase. DP instances within a DP-cell coordinates usage of the aggregate WAN bandwidth. DP-cells operate independent of each other.} In Fig.\ref{fig:design:dpsharing}(b), DP-1 and DP-2 form a single DP-cell. Each DP instance in a DP-cell is assigned a \textit{LocalDPRank} and aggregate WAN bandwidth is shared temporally between these DP instances based on their \textit{LocalDPRanks}. The bubbles between microbatches are because WAN communication is slower than compute. To eliminate such bubbles, we set number of DP pipelines in a DP cell as $C$, where $C$ is communication to compute ratio. 

\parab{($2$)} While scheduling forward passes for a microbatch on any DP pipeline, \db{\tname filters only those for which activations/gradients in memory at any point of time is within the peak memory limit.} 
As a result: ($a$) we are always within the peak memory limit, unlike Varuna, which may exceed memory limits, and ($b$) we don’t block computation and communication phases on other DP pipeline because of unnecessary utilization of the aggregate WAN bandwidth for transmitting activations/gradient that would have later resulted in exceeding peak memory limit anyway.
    
\parab{($3$)} \tname schedules the compute phase of a micro-batch $m$ in a DP pipeline at time $t$ only when its communication phase could be scheduled immediately next, without overlapping with the communication phase of \db{any other 
already generated schedule}. \db{In case of contention, \tname reschedules the compute phase for the micro-batch in a way that the ensuing communication phase does not overlap with any other network communication in the same DP-cell.}
As a result, DP-$2$ starts at time $1$ and DP-$1$ starts at time $5$ in Fig.\ref{fig:design:dpsharing}. 

\parab{($4$)} If a stage $k$ has both forward and backward tasks ready to be scheduled, \db{\tname 
prioritizes the backward pass to unlock processing at subsequent nodes}.

\subsection{DC selection}
\label{sec:design:dc}

\begin{table}[t] 
\centering 
\caption{Notations used in selecting GPUs for training.}
\label{tab:design:notation}
{
\small
\begin{tabular} {|c|c|}
\hline
\textbf{Notation} & \textbf{Explanation}\\
\hline
\hline
$D_{max}$ & Max. number of DP-cells \\
\hline
${Num\_GPU}$ & Number of GPUs in individual DCs (Map) \\
\hline
$C$ & Communication to compute ratio for PP \\
\hline
$P$ & Max. number of partitions \\
\hline
$total\_{time}$ & Total time for different DP-cells (Map) \\
\hline
$DCs$ & Ordered list of DCs \\
\hline
$Partitions$ & Partitions assigned to DCs (Map) \\ 
\hline
\end{tabular}
}
\end{table}

Above sections assume a given topology of GPUs and optimize for such topology. However, it is not always beneficial to use all the GPUs. For example, if $1$,$000$ GPUs are in the same DC, and $10$ GPUs are in a different DC, then it's better to forgo the $10$ GPUs and avoid slowdown in training time due to poor WAN conditions. 

\begin{figure*}[t] 
\vspace{-0.2in}
\begin{minipage}{1\linewidth}
\begin{algorithm}[H]
\caption{Algorithm to calculate latency for different DP-cells.}
\label{alg:design:algo}
\begin{flushleft}
        \textbf{INPUT:} $D_{max}, DCs, Num\_GPU, C, P$ \\
        \textbf{OUTPUT:} $total\_{time}$
\end{flushleft}
\begin{algorithmic}[1]
\For{$D$ in $\{1$ to $D_{max}\}$}
    \State $part\_{left} = P$;
    \For{$dc$ in $DCs$}
            \State $PP\_GPU = \lfloor\frac{Num\_GPU[dc]}{D \cdot C}\rfloor$;
            \State $part\_{assigned} = min(part\_{left}, PP\_GPU)$;
            \State $Partitions[dc] = part\_{assigned}$;
            \State $part\_{left} -= part\_{assigned}$;
            \If{$part\_{left}$ == 0}
                break;
            \EndIf        
    \EndFor
    \If{$part\_{left} > 0$}
        \State $PP\_time = \infty$;
    \Else
        \State $PP\_time = \texttt{get\_latency\_pp}(Partitions, D)$;
        \State $all\_reduce\_time = \texttt{get\_latency\_dp}(D \cdot C)$;        
    \EndIf
    \State $total\_{time}[D] = PP\_time + all\_reduce\_time$;
\EndFor
\end{algorithmic}
\end{algorithm}
\end{minipage}
\end{figure*}

\db{Now, we describe how \tname assists in finding the right number of GPUs from individual DCs. The goal here for \tname is to determine the right number of GPUs in individual DCs that maximizes training throughput (thus reduces training latency). Users can also run \textit{what-if} analyses to determine the impact of varying number of GPUs in multiple DCs, and choose} a set of GPUs to find a balance between the cost and the performance.

\textit{Our key observation is that we should (mostly) use all GPUs in a DC or none.} It is better to house all the GPUs in the smallest number of DCs possible to improve the training time, as the PP time is better when GPUs are in the same DC compared to having same number of GPUs in different DCs that requires communication over WAN. Thus, \db{\tname tries} to maximize number of GPUs used in the same DC and tries to minimize the number of DCs used.

Algorithm.~\ref{alg:design:algo} calculates the training latency, and the notations are shown in Table \ref{tab:design:notation}. The inputs to the algorithm include: ($a$) \db{an implicit ordering} of the DCs (e.g., based on cost of GPUs; \db{the default is based on decreasing order of GPU availability)}, ($b$) the number of available GPUs in each DC ($Num\_{GPU}$), ($c$) maximum number of DP-cells ($D_{max}$), ($d$) communication to compute ratio ($C$), and ($e$) number of partitions ($P$). $P$ is the ratio of total layers in a model to the number of layers fit on a model due to resources on GPU (e.g., GPU memory). Smaller partitions see smaller PP communication overhead. The output of the algorithm is the total time for different values of DP-cells ($D$). The users then decide $D$ depending on cost, performance, and other metrics. $D_{max}$ can be set to $\frac{\displaystyle{\sum Num\_{GPUs}}}{C \cdot P}$. At a high-level, the algorithm calculates the total training time of an iteration for each value (D) of DP-cell [1,$D_{max}$] that includes the time for running PP and all-reduce in DP. The compute time (including TP, if any) is constant across D that we ignore. 

To do so, the algorithm ($a$) first iterates DCs in an order based on cost, distance, or other metrics (line $3$), and then ($b$) calculates the number of GPUs for the PP stages as $\frac{{Num\_{GPU}[dc]}}{D \cdot C}$, as there are D DP-cells each with C individual DP pipelines (line $4$). It assigns more PP stages to DCs with more GPUs, ($c$) It assigns the number of partitions that is minimum of partitions left and number of GPUs in PP calculated in ($b$) (line $5$). ($d$) It stores the GPUs assigned in Partitions map (line $6$) and adjusts partitions left (line $7$). ($e$) This routine exits when all partitions are assigned or we run out of GPUs (line $8$-$12$). ($f$) Lastly, it calculates the total execution time for a given $D$ (line $14$-$17$). ($g$) \texttt{get\_latency\_pp} calculates the latency for temporal bandwidth sharing for a DP-cell. Similarly \texttt{get\_latency\_dp} calculates the latency for the all-reduce phase across DP pipelines.

\textbf{Calculating throughput:} As we show in \S\ref{sec:eval:selection}, such an algorithm then helps in finding the smallest $D$ (hence GPUs) that provides highest throughput. The throughput is calculated as $\frac{D \cdot C}{total\_{time}[D]}$. 

\rohan{\textbf{Discussion on selecting DCs}: Due to privacy regulations such as GDPR, it may not be possible to move data across jurisdictions. Many techniques can be used for compliance (\cite{gdpr:web}). Additionally, we can set enough DCs in such jurisdictions to consume the data locally.}

\textbf{Performance and cost modeling:} As mentioned above, we can plug-in any combination of DCs and number of GPUs in DCs and the above algorithm can calculate the best \textit{configuration} (DCs and set of GPUs) that provides highest throughput. We can plug-in different combinations and model the cost and performance trade-off \textit{without any deployment}. This way, we can choose the best configuration as per our needs and then deploy it on hardware.

\section{\iname}
\label{sec:idesign}

As seen already, \tname reduces the training time using \textit{temporal bandwidth sharing}. Additionally, it eliminates the bubbles between microbatches and collates them together. However, it does not eliminate the GPU compute bubbles where GPU is idle. We find that despite \tname finishing training faster (and improving GPU utilization), it results in up to  55\% time spent in bubbles.

Note that achieving 100\% GPU utilization (no bubbles) is challenging due to the inherent nature of dependencies between sequential stages of pipeline parallelism\cite{varuna:eurosys22, gpipe:neurips19, zb:arxiv23}. Training over WAN amplifies the bubbles due to longer latencies and smaller bandwidths on WAN. 

There are two strawman's approaches to reduce \db{such residual bubbles}, as follows. ($a$) \textit{Increasing compute}: as compute is shorter than communication in PP, one way is to increase compute to match communication. This can be accomplished by each GPU processing more layers. As GPU memory is limited, the layers could be loaded from CPU (like \cite{zero:sc21}). However, we find that such a design is bottlenecked by PCIe bandwidth, where the time to load a $1$ billion parameter layer from CPU is at least $100$~msec even after using pinned-memory \cite{pinnedmem:web} (larger than processing the layer itself). ($b$) \textit{Decreasing communication}: we tried reducing communication through compression that resulted in 2$\times$ slowdown in training latency while achieving similar loss rates (\S\ref{sec:eval:semantics}).

We take a novel approach to reduce bubbles. We build \iname to schedule a different workload during the bubbles. Specifically, we observe that the prefill phase of the inference workload is well suited for such scheduling. Inference requests have distinct prefill and decode phases\cite{splitwise:isca24, sarathi:osdi24} -- the prefill phase processes the input prompt and the decode phase generates the output tokens in an autoregressive fashion. The execution time of decode phase is unknown at the start as it depends on the \db{number of tokens} generated. In contrast, the execution time of prefill phase is highly predictable as it only depends on the input prompt (size) that is known when the request arrives. Leveraging this insight, we build \iname that provides \textit{prefill-as-a-service} that we detail next.

\subsection{\iname design}
\label{sec:idesign:design}

\iname runs prefill for already trained models such as LLama 3.1 while training a different model (e.g., GPT-5). The key goal is to schedule the prefill of inference requests without interfering with the training job. To do so, we do not simultaneously schedule training and prefill. Rather, the prefill kernels are issued only when the training job observes (predictable) bubbles. \rohan{This way there is \textit{no contention} on compute and memory bandwidth at GPUs. The only overhead comes from storing the models  (discuss later).} Based on the prefill estimated latency, we choose the GPU so that prefill ends \db{(compute becomes available again)} before the training job resumes (does not hamper training job). This approach \db{significantly improves the GPU utilization in training clusters}.

\begin{wrapfigure}{l}{0.55\textwidth}
\centering
\includegraphics[width=0.52\textwidth, page=1]{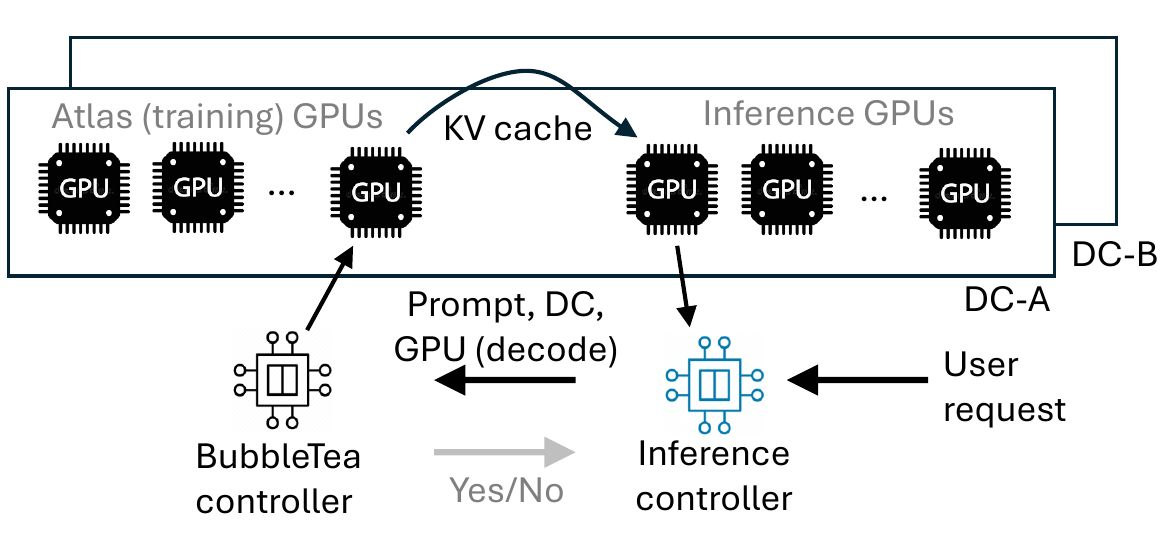}
\caption{\iname Architecture.}
\label{fig:idesign:arch}
\end{wrapfigure}

The prefill phase requires the model weights stored in the GPU memory. Let's assume that training uses model $M1$ (could be subset of the layers depending on the memory size), and inference uses model $M2$. One solution could be that we store both the models on CPU and load them to GPU memory before the respective jobs run. However, \db{we observed that} such dynamic loading incurs significant delays due to \db{limited} PCIe bandwidth. Assuming $M1$ and $M2$ each consume $60$~GB memory on a $80$~GB A$100$/H$100$ GPU, loading the layers with $64$~GBps PCIe bandwidth\cite{h100spec:web}\footnote{PCIe gen$5$ has bidirectional bandwidth of $128$~GBps; one-way bandwidth is $64$~GBps.} takes roughly $0.93$~second each. The GPU will be idle for this duration. 


Instead with \iname, we pre-store subset of layers of $M2$ on GPU, while we store $M1$ as-is. Consequently, \textit{$M2$ is sharded across more GPUs and we use PP \rohan{across different GPUs} to run $M2$}. This design choice incurs a small overhead in terms of memory capacity. E.g., for serving Llama$3$-$8$B  using 8 GPUs would require only $2$~GB memory (at fp$16$) per GPU. Transient activations and KV caches use the memory not in use in training. This way, \iname incurs very minor overhead. We do so to provide much larger memory for training model ($M1$) so that larger models such as GPT or Llama (405B) can be trained in geo-distributed settings.


\parab{\iname architecture:} Fig.\ref{fig:idesign:arch} shows the architecture in \iname. The user request is received by the \rohan{logically centralized} inference controller. The inference controller then chooses the DC (E.g., using locality; its logic is outside the scope of \iname), and sends the user prompt, DC and GPU details (for processing the decode) to the \iname controller. \iname controller then schedules the prefill on one of the \tname GPUs with enough capacity. After the completion of the prefill, the KV cache is transferred to the GPU specified by the inference controller for the decode phase (using Splitwise architecture\cite{splitwise:isca24}). The decode GPUs are in the same DC to use faster networks to transfer KV cache. In case \iname does not have enough capacity to process the prefill phase using training GPUs (e.g., the training is ongoing and there is no bubble at that time), it \db{immediately} informs the inference controller accordingly. 


\parab{Pipeline parallelism for prefill requests:} We do PP for inference model across GPUs. To do so, \iname forms a PP pipeline across GPUs of the same rank in individual DP-cells. Fig.\ref{fig:design:dpsharing}(b) shows a single DP-cell. Imagine another DP-cell with GPUs G-13 to G-24. GPUs G-1, G-13 and GPUs from subsequent DP-cells form a PP where each GPU is assigned a small number of layers. Note that, PP in \iname is formed only across GPUs in the same DC to reduce latency. As prefill latency is highly predictable, \iname schedules prefill only if there is available capacity. To do so, it finds first available PP pipeline that has bubble across GPUs to accommodate the prefill request scheduled using PP. Importantly, we find that doing PP is beneficial even if the entire inferencing model fits on the same GPU (\S\ref{sec:eval:iname2}) to reduce TTFT (Time-To-First-Token). Such PP improves the TTFT by 67\% for larger prefills (\S\ref{sec:eval:iname2}). We discuss this and \iname overhead in \S\ref{sec:eval:iname2}. This approach has no impact on TBT (Time-between-Tokens). The TTFT can be further reduced using \textit{chunked} prefills\cite{sarathi:osdi24} that we leave for future work. \rohan{No component becomes single point of failure. We discuss \iname controller overhead in \S\ref{sec:eval:iname}}.


\parab{Scheduling prefills:} \iname controller gets a rough plan for scheduling the microbatches from \tname and identifies the bubbles where prefill can be scheduled. It then waits for signals from the individual GPUs when they process microbatches (using hooks in PyTorch). It then assigns prefills during bubbles based on such signals. 

\parab{Scheduling decodes:} Once the prefill is computed on training GPUs, the KV cache is transferred to decode GPUs using existing designs such as Splitwise\cite{splitwise:isca24} and DistServe\cite{distserve:osdi24}. As such designs already decouple prefill and decode, they optimize for decodes independently (such as batching). \iname uses such designs for scheduling decodes.

\parab{LMs where \iname helps:} We recommend running smaller LMs including small Llama\cite{llama:web}, Phi\cite{phi:web}, Mistral\cite{mistral:web} models. Recall that the training model in \tname and inference model in \iname are to be stored in the GPU memory. We do not use them simultaneously, so there is no contention on memory bandwidth and compute SMs on GPU. To keep the memory overhead of the inference model small, we only recommend using the small LMs for inference. This way most of the memory is available for the training model and larger models such as GPT or Llama (405B) can be trained in geo-distributed settings. \S\ref{sec:eval:iname2} shows the overhead in \iname.

Lastly, while we propose interleaving prefill (from inference workloads) with the training, there could be other workloads that can be scheduled during bubbles. We detail it in \S\ref{sec:related}. 

\section{Evaluation}
\label{sec:eval}

We evaluate \tname and \iname using testbed experiments and large-scale simulations. 
\begin{wrapfigure}{l}{0.55\textwidth}
\centering
    \begin{minipage}{.55\textwidth}
        \centering
        \subfigure[GPT-A (L,H = 4K, 4K)]
        {
        \includegraphics[width = 0.95\textwidth, page=1]{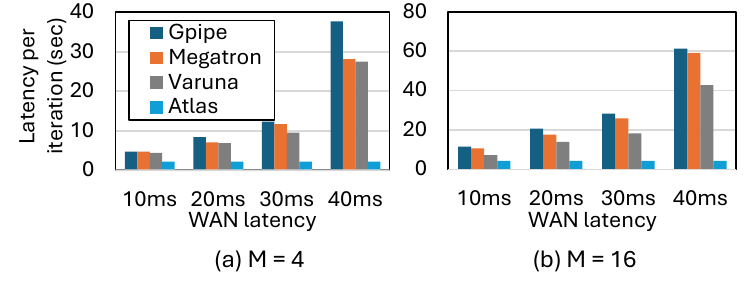}
        \label{fig:eval:singleconn:1}
        }
        \\
        \subfigure[GPT-B (L,H = 6K, 8K)]
        {
        \includegraphics[width = 0.95\textwidth, page=2]{figs/testbed-multi-conn.pdf}
        \label{fig:eval:singleconn:2}
        }
        \vspace{-0.1in}
        \caption{[Testbed] Training latency  (for one iteration) for GPT-A and GPT-B for different WAN latencies and number of microbatches (M).}
        \protect\label{fig:eval:singleconn}
    \end{minipage}%
    \\ 
    \begin{minipage}{0.55\textwidth}
        \centering
        \subfigure[GPT-A (L,H = 4K, 4K)]
        {
        \includegraphics[width = 0.75\textwidth, page=3]{figs/testbed-multi-conn.pdf}
        \label{fig:eval:bw:1}
        }
        \\
        \subfigure[GPT-B (L,H = 6K, 8K)]
        {
        \includegraphics[width = 0.75\textwidth, page=4]{figs/testbed-multi-conn.pdf}
        \label{fig:eval:bw:2}
        }
        \vspace{-0.1in}
        \caption{[Testbed] Iteration time for GPT-A and GPT-B across different number of microbatches (M).}
        \protect\label{fig:eval:bw}
    \end{minipage}%
\end{wrapfigure}

The key observations are: ($a$) \tname, with its optimizations, improves the cross-DC training time by up to $17${}$\times$ compared to baselines like Varuna, GPipe and Megatron. ($b$) Even when the baselines leverage multiple TCP connections, \tname improves training time by up to $1.82${}$\times$ than baselines using temporal bandwidth sharing. ($c$) \tname is able to scale the throughput as we add more DCs. This is in stark contrast with findings in \S\ref{sec:limitations} where throughput reduced when using WAN. ($d$) \iname could schedule the prefill requests during bubbles and achieves a GPU utilization of up to $94\%$ in the training cluster, up from $45\%$ when using only \tname.



\subsection{Setup}
\label{sec:eval:setup}

\parab{Setup for testbed experiments:} Our evaluation uses a setup similar to Fig.\ref{fig:limit:setup} consisting of $12$~GPUs in $3$~DCs ($4\times3$). We vary the latency between DCs using \texttt{tc} command. We set the maximum cross-DC bandwidth between two nodes to 5 Gbps (\S\ref{sec:design:tcp}), whereas bandwidth between two nodes in the same DC is capped to 100 Gbps.


All GPUs are A$100$ $80$~GB running PyTorch. We compare the training time using \tname versus baselines including GPipe\cite{gpipe:neurips19}, Megatron\cite{megatron:arxiv19} and Varuna\cite{varuna:eurosys22}. For the baselines, we divide the GPUs among $3$~DP pipelines with $4$~PP stages. For \tname, we have $1$~DP-cell with $3$~DP pipelines with $4$ PP stages as above. While we do not leverage TP \db{for the demonstrations, TP could be used intra-DC}. We use GPT-A and GPT-B models (\S\ref{sec:limitations}) for the evaluation. As mentioned in \S\ref{sec:limitations}, we set the number of layers to fit on $4$~PP stages. However, individual layer sizes in GPT-B model are higher than Llama 3 80B model. We set the number of microbatches to $4$ (inline with the $4$ PP stages) and $16$.

\subsection{\tname improves training time}
\label{sec:eval:training}

We show that \textit{\tname is able to improve the training time by up to $17${}$\times$ compared to the baselines using the $2$ key ideas of using multiple TCP connections and temporal bandwidth sharing}. We compare \tname against default baselines (that do not use multiple connections; we give the benefit of multiple TCP connections to the baselines in the next section). We vary the WAN latency from $10$~msec to $40$~msec. Note that, irrespective of the latency, \tname gets same bandwidth due to multiple TCP connections. In contrast, bandwidth for the baselines reduces as WAN latency increases (Table \ref{tab:mot:bw}).

As shown in Fig.\ref{fig:eval:singleconn}, \tname is able to reduce the training time substantially. We find: (a) Compared to Gpipe, Megatron and Varuna, \tname is able to reduce the training time by up to $17${}$\times$, $13${}$\times$, and $12${}$\times$ resp. across different WAN latencies and number of microbatches (M). (b) As expected, \tname gains increase as WAN latencies increase. This is because the baselines get lower bandwidth for single TCP connection as WAN latencies increase. That said, even for a small WAN latency of $10$~msec, \tname is able to improve the training time by up to $2.68${}$\times$. (c) the gains also reduce for M$=16$ compared to M$=4$. This is mainly because the compute takes higher fraction of the total latency for M$=16$. (d) the gains in \tname also reduce for GPT-B compared to GPT-A, again due to compute taking higher fraction of total latency.

\parab{Temporal bandwidth sharing is crucial}
\label{sec:eval:tempo}
In this experiment, \textit{we show that just using temporal bandwidth sharing can still substantially improve the training time (by up to $1.82${}$\times$)}. We continue to use the $12$~GPU cluster from previous section. In this experiment, we give the benefit of using multiple TCP connections to all the baselines too alongside \tname. This way, we primarily evaluate the benefits of temporal bandwidth sharing (\S\ref{sec:design:jct}). Fig.\ref{fig:eval:bw} shows the training latency per iteration (average) for GPT-A and GPT-B. We also measure such latencies for $4$ and $16$ microbatches. We found that \tname improves the training latency by up to $1.82${}$\times$, $1.72${}$\times$ and $1.52${}$\times$ compared to Gpipe, Megatron and Varuna. Compared to Varuna, we found that \tname is able to start the microbatches sooner at the GPUs higher in the PP pipelines that in turn starts the backward passes sooner.

Together, \tname achieves GPU utilization to near 45\%.

\subsection{Efficient use of cross-DC GPUs}
\label{sec:eval:multidcs}

\begin{figure*}[t]
\centering
\includegraphics[width=0.75\textwidth, page=2]{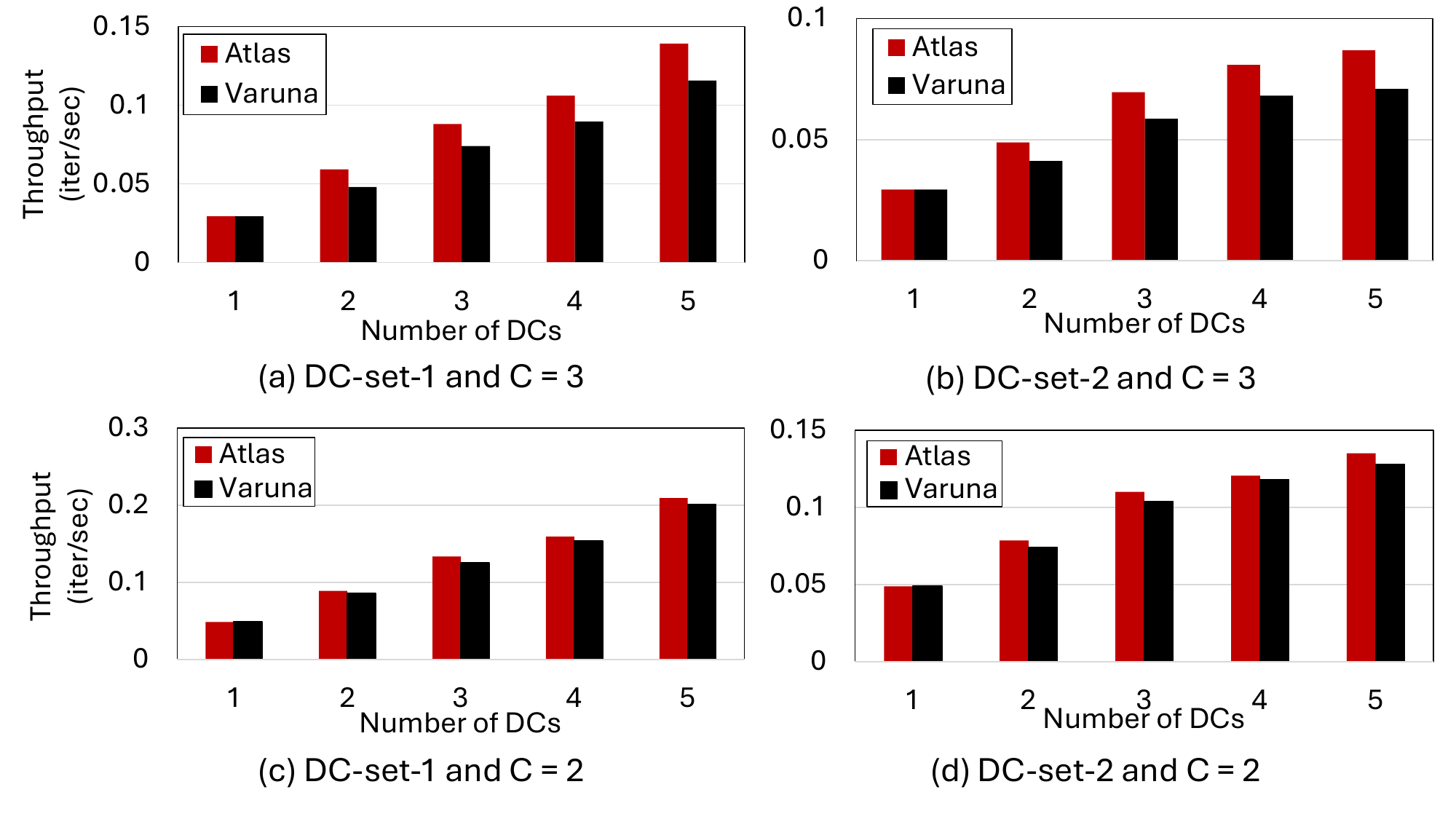}
\caption{[Simulation] Throughput in \tname and Varuna for different DC-sets and communication to compute ratio (C).}
\label{fig:eval:sim}
\end{figure*}

\textit{We demonstrate now through simulations how \tname improves throughput over Varuna by up to $48\%$ (potentially saving $10$s of millions of dollars at scale) by using GPUs across different DCs.} In the previous experiments, we used $12$ GPUs. In this section, we use simulations to evaluate the benefits on \tname over Varuna for bigger clusters.

In this experiment, we consider two DC-sets (simulated). DC-set-1 contains $600$~GPUs in each DC and we vary the number of DCs from $1$ to $5$. DC-set-2 contains GPUs, likewise, in $5$~DCs as [$600$, $500$, $400$, $300$, $200$]. In this experiment, we give the benefits of using multiple TCP connections to both \tname and Varuna. For both \tname and Varuna, we assume $5$~Gbps bandwidth is available between 2 nodes in different DCs irrespective of the latency. This way, we mainly focus on the benefits due to temporal bandwidth sharing in \tname. As mentioned in \S\ref{sec:design:3d}, despite multiple TCP connections, communication still takes 3-4$\times$ compute latency. In this experiment. we consider two communication to compute ratios (C) as $4\times$ and $2\times$. We set number of layers (microbatches and PP degree) to $60$. In Varuna, we set the number of DP pipelines as per the available GPUs and PP degree (60). In \tname, we set the number of DP pipelines calculated using Algorithm.~\ref{alg:design:algo} (\S\ref{sec:design:dc}). However, we found that such an algorithm used all GPUs. For DC-set-2, it assigned more PP stages to DCs with more GPUs. In \tname, we set the number of DP pipelines in a DP-cell to $C$ as that's the minimum number of DP pipelines required to overlap communication time with compute.

\textbf{Scalability with increasing DCs:} Fig.\ref{fig:eval:sim} shows the \db{achieved training throughput (iterations/second)} in $4$ cases. It can be seen that training throughput improves with more GPUs. \db{With \tname, for [DC-set-$1$, C$=4$, $5$~DCs], we get roughly $4.7${}$\times$ throughput compared to using a single DC.} Similarly for [DC-set-$1$, C$=2$, $5$~DCs] the speedup is roughly $4.3${}$\times$. Similar trends occur for Varuna too. This shows that the multiple forms of parallelism used in \tname, where PP is used across DCs, \db{is effective in using GPUs from different DCs}.

Next, we consider the benefits of using \tname over Varuna. For C $=4$, where communication time is $4${}$\times$ compute time, \tname substantially outperforms Varuna. For both DC-set-$1$ and DC-set-$2$, the improvement in throughput over Varuna is up to $48\%$. The speedup is mainly because, \db{with temporal bandwidth sharing, \tname is able to improve the network utilization} and start the forward and backward passes sooner at nodes higher up in pipelines. However, the benefits shrink for C$=2$, \db{along the lines of expectation}, where \tname improves throughput by up to $25\%$ because, at lower C, there are already fewer bubbles and \tname has less opportunities to pack the individual passes.

\subsection{Cross-DC GPU balancing}
\label{sec:eval:selection}

\begin{wrapfigure}{l}{0.7\textwidth} 
\centering
\includegraphics[width=0.57\textwidth, page=1]{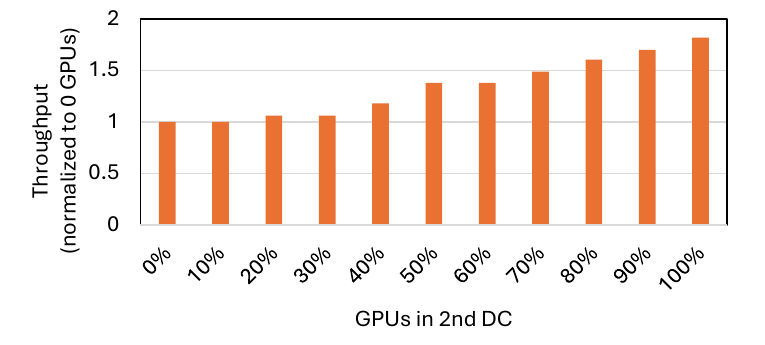}
\caption{[Simulation] Throughput in \tname using 2 DCs.}
\label{fig:eval:selection}
\end{wrapfigure}

In the last experiment, we used all the GPUs available. In this experiment, we show that \textit{it's not always better to use all available GPUs \db{across DCs}, and Algorithm \ref{alg:design:algo} is needed to calculate the smallest number of GPUs required for the highest throughput.} In this experiment, we pick two DCs -- the first DC has a fixed number ($600$) of GPUs and we vary the number of GPUs in the second DC from $0$ to $600$ in steps of $10\%$ (denoted by F). Through simulations (same setup as last experiment with C$=2$), we \db{calculate the training throughput}. 
We use Algorithm.~\ref{alg:design:algo} (\S\ref{sec:design:dc}) to calculate the optimal number of GPUs in individual DCs for highest throughput.

Fig.\ref{fig:eval:selection} shows the improvement in training throughput normalized to using GPUs in a single DC. It can be seen that when F increases to $10\%$, there is no improvement in throughput. \db{Algorithm.~\ref{alg:design:algo}, in this case, falls back to only first DC (no GPUs from second DC).} This is because, \db{for such F}, we need to send the activations from first DC to second DC over the WAN. \db{This inflates the latency to complete one iteration by roughly $11\%$ and, hence, erases any throughput gains by adding the additional $10\%$ GPU compute.} We observe similar behavior when increasing $F$ from 20\% to 30\%. This \db{reinstates the} need for Algorithm.~\ref{alg:design:algo} that does a sweep across different options and chooses the one that optimizes for a given metric (cost, latency, throughput, etc.). 

Lastly, Algorithm.~\ref{alg:design:algo} is fast -- even for $5$ DCs with $600$ GPUs each, it took $55$~seconds to execute on a single CPU machine. Engineers can  quickly do exhaustive \textit{what-if} analysis and estimate the impact of different number of DCs and GPUs.

\subsection{\iname boosts GPU utilization}
\label{sec:eval:iname}

In this section, we show how \textit{\iname can significantly improve the GPU utilization by scheduling prefill stages of inference requests during bubbles in \tname}. We use the $12$ GPU setup and schedule prefills as mentioned in \S\ref{sec:idesign:design} running GPT-A. As there is only one DP-cell, we set the PP depth of the inference model to $1$. Recall that the \iname controller combines ($1$) the scheduling plan from \tname controller and ($2$) signals from the GPUs when they finish processing training microbatches. Accordingly, \iname issues the inference kernels.

\begin{wrapfigure}{l}{0.7\textwidth} 
\centering
    \begin{minipage}{.6\textwidth}
        \centering
        \includegraphics[width=0.92\textwidth, page=1]{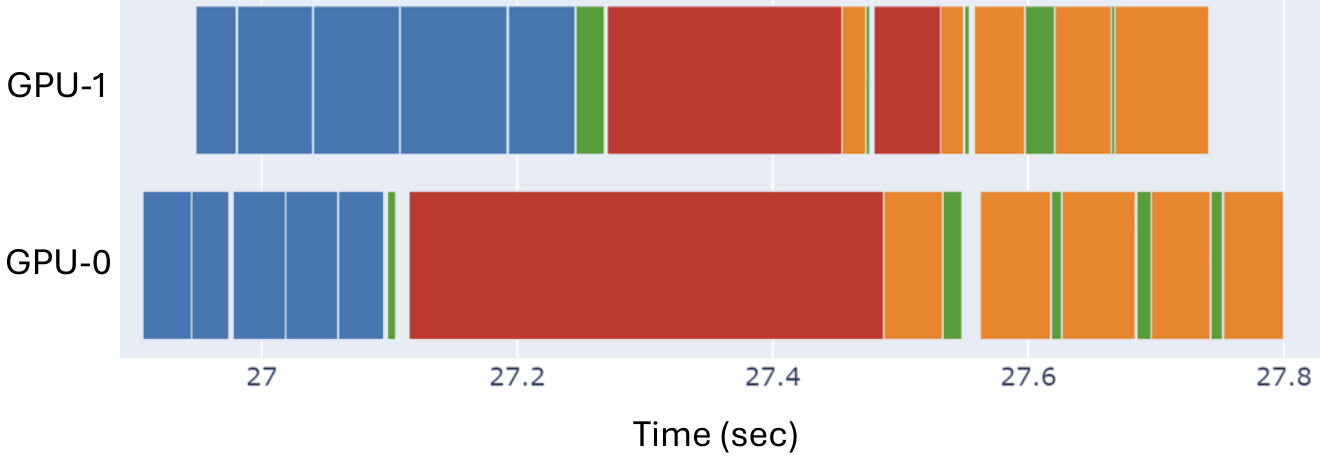}
        \caption{[Testbed] Execution of forward (blue), backward (orange), recompute (green) and \textit{inference prefill} (red) on two GPUs with \iname and \tname.}
        \label{fig:eval:iname:share}
    \end{minipage}%
    \\ 
    \begin{minipage}{0.6\textwidth}
        \centering
        \includegraphics[width=0.8\textwidth, page=1]{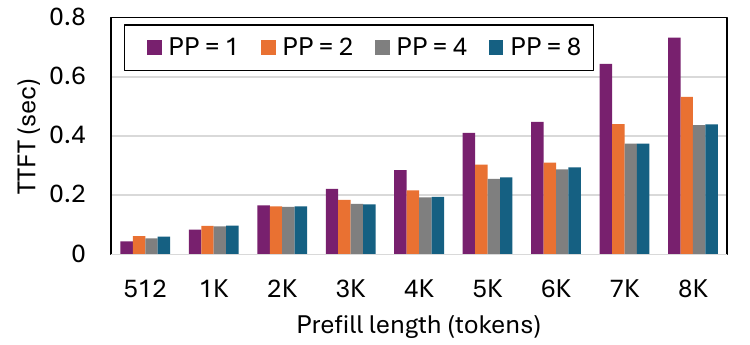}
        \caption{[Testbed] TTFT in using different PP degree in \iname for Llama$3$ $8$-billion.}
        \label{fig:eval:iname:pp}
    \end{minipage}%
\vspace{-0.1in}
\end{wrapfigure}

Fig.\ref{fig:eval:iname:share} shows the GPU utilization for two of the GPUs in a single (training) pipeline. We make the following observations: (a) \iname is able to schedule the prefill stages perfectly during the training bubbles. This shows that \iname controller can accurately detect the bubbles (through combining scheduling plan and GPU signals). (b) In doing so, \iname boosts the GPU utilization to around $94\%$ leaving very small residual bubbles. (c) There is no impact on the training job. We found \iname leaves some bubbles between prefill and training microbatches that helps resume training without delay. (d) \iname does not schedule prefill if the bubble is not big enough.

\parab{Impact on GPU utilization:} Fig.\ref{fig:eval:iname:share} shows \iname goes one step further than \tname and improves the GPU utilization from $45\%$ (\tname-only) to $94\%$ using inference prefills. 


\parab{\iname overhead:} In the 12-GPU experiments, we could find the prefill bubble to schedule inference request in $<$100$\mu$sec. We also do a large scale simulation over 1000 GPUs in a single DC. The GPUs are divided across 50 DP-cells. We replay inference workload as coding dataset from \cite{coding:trace:web}. We found that \iname is able to find the training bubble within 200 $\mu$sec and the wait time (queuing delay) is within 8 msec.

\subsection{\iname PP overhead}
\label{sec:eval:iname2}

\rohan{Recall that, to save the memory overhead of storing inference models in \iname, we split the models across different GPUs and use PP. This only impacts TTFT and not TBT as decode runs on dedicated GPUs (\S\ref{sec:idesign:design}). In this experiment, we use a DGX node using $8$ A100 GPUs connected via NVLink. We use the Llama$3$ $8$ billion model. We vary the PP degree from $1$ to $8$ and set the number of GPUs to the PP degree. Fig.\ref{fig:eval:iname:pp} shows the TTFT for different PP degrees as we vary the prefill length. We make two key observations: (a) Increasing the PP degree negatively impacts the TTFT when prefill length is small. For $512$ tokens, PP of degree $8$ inflates TTFT by $29\%$. However, in absolute numbers, the increase is just $16$~msec and the TTFT is well under SLOs\cite{distserve:osdi24}. The increase is primarily due to communication overhead in PP. (b) As the prefill length increases, the TTFT improves with higher PP degree. In fact, for higher prefill lengths, higher PP degrees actually perform better than lower PP degrees. E.g., for prefill length of $8K$ tokens, the TTFT for PP=$1$ is $67\%$ higher than PP=$8$. We observe that, for larger prefills, the compute becomes saturated even for a subset of the layers. Consequently, the weights are to be swapped in and out, resulting in higher latency. With PP=$8$, the layers are already partitioned and there is no need to swap, thus improving TTFT. At PP=$8$, each GPU only uses (small) $2$~GB memory for the inferencing model.}   

\rohan{
\textbf{LMs where \iname helps:} As noted previously, we currently limit the memory available to \textit{inference models} to $1$-$2$~GB so that most of the GPU memory is available to the training model (such as GPT or Llama-405B). \iname can still support many small/medium LMs from the Llama, Phi, Mistral, and Gemma family for inference. As we relax this constraint, we can support more models that we leave for future work. Additionally, \iname can easily support batch (latency insensitive) requests\cite{distserve:osdi24}. Note that, we only need to use smaller models for inference (in \iname). In \tname, we focus on larger LMs such as GPT or LLlama-405 for training.
}

\subsection{Comparing with semantics altering methods}
\label{sec:eval:semantics}

\tname does not change the semantics (functionalities) of the protocols -- it runs \textit{standard} PP and DP to maintain the accuracy. \db{While we tried other designs that change semantics to improve training time at the cost of lower accuracy, our engineers suggest against such coercive approaches.}

We tried to reduce the communication time in PP through compression techniques from \cite{ppcompress:mlsys24} and SVD~\cite{svd:web}. They achieve good compression but suffer from \db{model accuracy loss and/or training time inflation}. \cite{ppcompress:mlsys24} also notes similar observation for many compression techniques such as Top-K and quantization. Additionally, we found that SVD based compression takes roughly $2${}$\times$ compute time to achieve the same loss rate without using compression. For such reasons, we do not use methods that alter the semantics of the protocols. That said, key ideas in \tname are complimentary to semantic altering methods. Using multiple TCP connections and temporal bandwidth sharing can work in tandem and can improve training latency.

\section{Related work}
\label{sec:related}

\tname and \iname together improve the training time and GPU utilization when using GPUs from different DCs. We detail the related work below: 

\parab{Geo-distributed training}: There have been much work on geo-distributed training. \cite{gaia:nsdi17} optimizes for parameter server based training that is orthogonal to \tname and does not use PP or DP like parallelism. \cite{geo:sigcomm24, fusionai:arxiv23} train for consumer held devices. \cite{geo:sigcomm24} optimizes for all-reduce traffic. However, they change the semantics that can affect the model accuracy. Similarly, works change the semantics either through quantization\cite{quant1:neuro, quant2:arxiv, quant3:neurips17} or sparsification\cite{sparsification1:neurips17, sparsification2:arxiv22, sparsification3:mlsys21}, or a combination of both\cite{cocktainsgd:icml23}. Existing works use different strategies for communication frequencies\cite{localsgd1:neurips19, localsgd2:aaai19, localsgd3:neurips21} that again can affect the model accuracy. In contrast, \tname does not change the semantics preserving the accuracy. Like \tname, \cite{decentralized:neurips22} structures the DP and PP among \textit{heterogeneous} nodes. However, they assign tasks to \textit{all} nodes that may inflate training time. Additionally, such works do not take advantage of DC-like settings that are more homogeneous. \cite{mast:osdi24} focuses on assigning training jobs to individual DCs and do not focus on speeding up training or GPU utilization by making use of GPUs in multiple DCs. Lastly, \cite{sailor:arxiv25} focuses on searching best configurations that \tname can leverage. Similarly, \tname is not tied to any one training schedule and  can work with schedule computed by other systems such as CrossPipe\cite{crosspipe:arxiv25}. \tname compliment such systems using temporal bandwidth sharing and using multiple TCP connections.

\parab{Mixing training and inference workloads}: Like \iname and \tname, existing works have also looked at combining training and inference workloads. Orion\cite{interference:eurosys24} looks at fine grained scheduling of such workloads to reduce interference. Usher\cite{usher:osdi24} focuses on interference due to different inference models. Sarathi\cite{sarathi:osdi24} mix prefill and decode phases of inference. \cite{inference:arxiv} shows significant interference when scheduling prefills and decodes together. \cite{pipefill:arxiv24} uses other workloads to fill in bubbles. We focus on decoupling training and inference to avoid interference. Lastly, Nvidia supports multiple instances running concurrently on a GPU through MIG\cite{mig:web}. We do not use it as we do not run training and inference concurrently. In fact, we increase the GPU utilization by repurposing idle GPUs for inference.

\parab{Improving network communication for training:} Recent works have also focused on improving the communication for training workloads. \cite{metatraining:sigcomm24} improves RDMA over Ethernet for training. NCCL\cite{nccl:web}, TACCL\cite{taccl:nsdi23}, TECCL\cite{teccl:sigcomm24} and RCCL\cite{rccl:web} optimize communication when GPUs are in the same DC, and do not optimize for TCP used for WAN communication. \db{CASSINI~\cite{cassini:nsdi24} explores traffic interleaving for multiple training jobs while our work improves communication for individual large training jobs in a cross-DC WAN setting leveraging temporal bandwidth sharing as one of the key techniques. Lastly, using multiple TCP connections have been used previously including \cite{mptcp1, mptcp2, mptcp3}}.

\parab{Reducing pipeline bubbles:} Works such as ZB\cite{zb:arxiv23} try to eliminate bubbles during forward and backward passes. However, such works: (a) continue to suffer from bubbles due to lower WAN bandwidth\cite{crosspipe:arxiv25}, and (b) require higher memory. Such a design is complimentary to temporal bandwidth sharing design in \tname. Works such as Hydro\cite{hydro:osdi23} use it to try out its trial jobs (not including inference). PipeFischer\cite{pipefisher:mlsys23} uses it to schedule parts of the existing job such as scheduling refresh the curvature and inverse matrices. PP in \tname is complimentary to PP in Megascale\cite{megascale:nsdi24} that can further benefit \tname.
\section{Conclusion}
\label{sec:conc}

The widespread adoption of language models (LMs) has caused a huge surge in demand for GPUs. Training large LMs is important to improve acccuracy and support larger context lengths. However, training large LMs is increasingly getting challenging as such training requires a large GPU fleet that's challenging to house in a single data center (DC). In this paper, we focus on training large LMs when GPUs are distributed across different DCs. We build \tname and \iname as complimentary tools to improve the training time and GPU utilization. \tname improves the training time using multiple design choices including novel temporal bandwidth sharing, but still results in bubbles (idle GPU time). \iname schedules prefill phase of inference requests during bubbles to substantially improve GPU utilization. Together, \tname and \iname improve the training time up to $17${}$\times$ and achieve GPU utilization of up to $94\%$.

\bibliographystyle{abbrv}
\bibliography{references}

\begin{appendices}

\section{\tname lessons for MoE training}
\label{sec:app:moe}

In this paper, we considered training for dense models with all parameters active at a time. However, there are Mixture-of-Expert (MoE) models that route tokens to a subset of experts, reducing number of active parameters at a time. Like TP, the bandwidth heavy all-reduce across different experts is in the critical path\cite{moe:atc23, moe:ppopp22}. Consequently, \iname architecture could be well suited for MoE model training where, like TP, EP (Expert Parallelism) is done within the DC and PP is done across DCs. The key lessons from \tname are also applicable to MoE training. E.g., MoE training will benefit from speeding WAN transfer latency using multiple TCP connections as well as temporal bandwidth sharing. Additionally, we can improve the GPU utilization using \iname. We leave this for future work.
\end{appendices}

\end{document}